\documentclass{elsart5p}

\usepackage{graphicx}

\usepackage{amssymb}

\begin{document}

\begin{frontmatter}



\title{Higher order terms in the geometric resonance of open orbits in unidirectional lateral superlattices}


\author{Akira Endo\corauthref{cor1}},
\author{Yasuhiro Iye}

\address{Institute for Solid State Physics, University of Tokyo,\\
5-1-5 Kashiwanoha, Kashiwa-shi, Chiba, 277-8581 Japan}

\corauth[cor1]{Corresponding author. E-mail: akrendo@issp.u-tokyo.ac.jp}

\begin{abstract}
The geometric resonance of open orbits in unidirectional lateral superlattices has been examined with high magnetic-field resolution. Magnetoresistance oscillations periodic in $1/B$, analogous to the well-known commensurability oscillations but orders of magnitude smaller both in magnitude and in the magnetic-field scale, have been observed superposed on the low-field positive magnetoresistance. The periodicity in $1/B$ can be interpreted in terms of higher order resonances.
\end{abstract}

\end{frontmatter}

\begin{figure}
\includegraphics[bbllx=5,bblly=30,bburx=595,bbury=515,width=8.5cm]{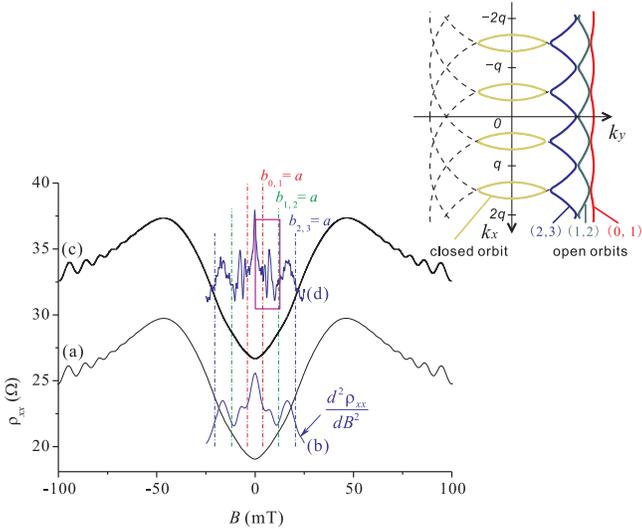}%
\caption{(a) Magnetoresistance trace $\rho_{xx} (B)$ of a ULSL taken by standard resistance measurement. (b) The second derivative $d^2 \rho_{xx}/dB^2$ obtained by numerically differentiating (a) by $B$ twice. (c) Magnetoresistance $\rho_{xx}(B)$ reproduced by numerically integrating the experimentally acquired first derivative $d\rho_{xx}/dB$ shown in Fig.\ \ref{rawdRdB}. The vertical scale is adjusted so as to present the same magnitude of the positive magnetoresistance as in (a), and the trace is offset for clarity. (d) The second derivative $d^2 \rho_{xx}/dB^2$ deduced by the numerical differentiation of $d\rho_{xx}/dB$ in Fig.\ \ref{rawdRdB}. The box indicates the portion to be replotted in Fig.\ \ref{d2dBBBI}(a). The vertical dot-dashed lines mark the locations of the occurrence of the fundamental geometric resonance, $n$=1 in Eq. (\ref{GROO}). The upper-right inset depicts the open and closed orbits. \label{magres}}
\end{figure}

Two-dimensional electron gases (2DEGs) subjected to one-dimensional periodic modulation, namely the unidirectional lateral superlattices (ULSLs), exhibit a wealth of intriguing magnetotransport phenomena in response to intricately combined effect of the modulated potential and the magnetic field $B$ applied perpendicular to the 2DEG plane. Two renowned examples are the low-field positive magnetoresistance (PMR) emanating from $B$=0 \cite{Beton90P} and the commensurability oscillations (CO) originating from the commensurability between the period $a$ of the modulation and the cyclotron radius $R_\mathrm{c}$=$\hbar k_\mathrm{F}/e|B|$, with $k_\mathrm{F}$=$\sqrt{2 \pi n_e}$ the Fermi wave number and $n_e$ the electron density \cite{Weiss89}. Both PMR and CO are basically semiclassical phenomena that can be explained by the semiclassical motion of electrons drifting around the modulated potential landscape \cite{Beton90P,Endo05P,Beenakker89}. Another class of interesting aspect in superlattices arises from the formation of the minibands and minigaps, or equivalently, from the diffraction (Bragg reflection) by the periodicity whose length-scale is orders of magnitude larger than that of the host crystals. Due to the diffraction, cyclotron orbits are reconstructed into sets of open and closed orbits that are composed of sections of arcs cut out from the cyclotron orbits (see the inset of Fig.\ \ref{magres}). Two types of magnetotransport phenomena associated with such reconstructed orbits are known to date: the quantum oscillations of closed orbits \cite{Deutschmann01} and the geometric resonance of the open orbits \cite{Endo05N}. The latter, to be denoted henceforth by an acronym GROO, has recently been uncovered by the present authors. It is observed as small amplitude oscillations superposed on the PMR and interpreted as resulting from the commensurability between the width $b_{j,k}$ of the open orbits [see Eq.\ (\ref{OOwidth}) below] and the period $a$ of the ULSL\@. In previous publications \cite{Endo05N,Endo05I,Endo06E,Endo07I}, we showed only the lowest and the second order resonances in which $b_{j,k}$ equals $a$ and $2a$ [$n$=1 and 2 in Eq.\ (\ref{GROO})], respectively. These are obviously not enough for the experimental verification of the expected periodicity in $1/B$ to be addressed below. In the present paper, we report the observation of higher order resonances (up to the fifth order), which is brought to light by the improvement both in the magnetic-field resolution and in the signal-to-noise (s/n) ratio in the resistance measurements. The oscillations bear striking similarity with the CO observed at higher magnetic fields, albeit with orders of magnitude reduced scales. 

\begin{figure}
\includegraphics[bbllx=50,bblly=50,bburx=520,bbury=400,width=8.5cm]{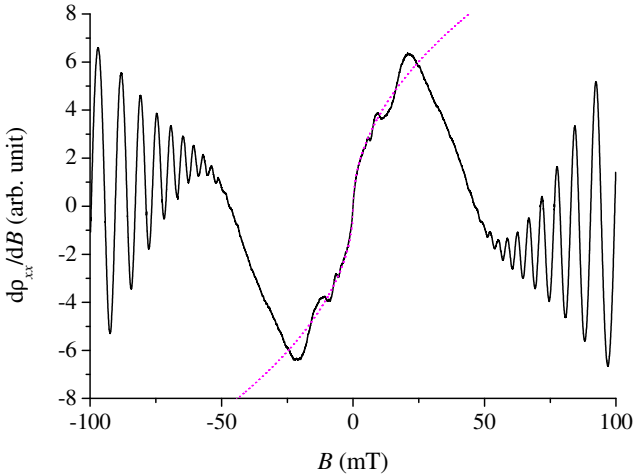}%
\caption{$d\rho_{xx}/dB$ directly measured by the double lock-in technique described in the text. The function $\alpha$ $\mathrm{sgn} (B) \sqrt{|B|}$ ($\alpha$=38.2 T$^{-1/2}$ in the present unit) is shown by a thin dotted curve, which approximately describes the slowly-varying background of $d\rho_{xx}/dB$ for $|B| <$ $\sim$25 mT\@. \label{rawdRdB}}
\end{figure}

The sample and the experimental techniques employed in the present study are basically the same as those described in the previous publications \cite{Endo05N,Endo05I,Endo06E,Endo07I}. An electrostatic potential modulation with the period $a$=184 nm is introduced to a conventional GaAs/AlGaAs 2DEG (the mobility $\mu$= 105 m$^2$V$^{-1}$s$^{-1}$ and the electron density $n_e$=3.0$\times$10$^{15}$ m$^{-2}$ obtained after LED illumination) exploiting the piezoelectric effect induced by the strain generated by a grating of electron-beam resist placed on the surface. The resultant periodic modulation seen by the electrons, $V(x)$=$\sum_j {V_j} \cos (j q x)$ with $q$=$2\pi/a$, can be evaluated by detailed Fourier analyses of the CO \cite{Endo05HH,Endo06E,Endo08FCO}; up to the fourth harmonics are detected for the present sample with $V_1$=0.31 meV, $V_2$=0.10 meV, $V_3$=0.07 meV, and $V_4$=0.05 meV\@.

\begin{figure}
\includegraphics[bbllx=30,bblly=50,bburx=510,bbury=420,width=8.5cm]{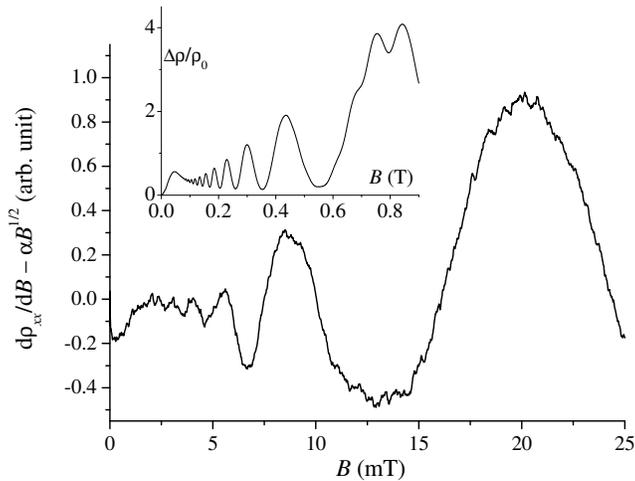}%
\caption{Oscillatory part of $d\rho_{xx}/dB$ for $B>0$ obtained by subtracting the slowly-varying part $\alpha \sqrt{B}$. The inset shows the commensurability oscillations of the same sample for comparison. Note the similarity between the two traces, albeit with orders of magnitude difference in the horizontal scale. \label{subtract}}
\end{figure}

A magnetoresistance trace measured by a standard low-frequency (70 Hz) lock-in technique at 4.2 K is shown in Fig.\ \ref{magres}(a), and also in the inset of Fig.\ \ref{subtract} up to higher magnetic fields. A simple way to extract small amplitude oscillations from the magnetoresistance trace is to take the derivative with respect to $B$. The differentiation may be done numerically as in Fig.\ \ref{magres}(b), which shows the second derivative $d^2 \rho_{xx}/d B^2$ obtained by numerically differentiating the trace in Fig.\ \ref{magres}(a) twice. For the numerical differentiation to be performed, however, it is usually necessary to take a measure to smoothen the data beforehand, which entails a certain degree of deterioration in the magnetic-field resolution. An alternative procedure, more appropriate for higher resolution and sensitivity measurement, is the direct acquisition of the first derivative $d \rho_{xx}/dB$ by the double lock-in technique \cite{Endo05N}. In the technique, a small amplitude ac modulation magnetic field ($B_\mathrm{mod}$=2 mT, 3.5 Hz) is superimposed on the dc magnetic field sweep, and the component of the resistance (output of the lock-in amplifier) that follows the $B_\mathrm{mod}$ is recorded by the second lock-in amplifier. The first derivative $d \rho_{xx}/dB$ thus acquired at 4.2 K is shown in Fig.\ \ref{rawdRdB}. In order to meet the requirement of the high magnetic-field resolution for the present purpose, we adopted slow sweep rate of the dc magnetic field ($dB/dt$=0.04 mT/s) with the data taken at every 0.01 mT\@. Although the magnetic-field resolution cannot be much better than $B_\mathrm{mod}$, such a high data acquisition rate is beneficial for obtaining good statistics of the data and thus for minimizing the deterioration in the magnetic-field resolution mentioned above during the course of further numerical differentiation. In order to further improve the s/n ratio, we take the average of sets of data taken by several magnetic-field sweeps; in Fig.\ \ref{rawdRdB}, the average of six sweeps is shown. In Figs.\ \ref{magres}(c) and (d), we plot $\rho_{xx}$ and $d^2 \rho_{xx}/dB^2$ obtained by numerically integrating and differentiating the $d\rho_{xx}/dB$ shown in Fig.\ \ref{rawdRdB}, respectively. The traces basically reproduce those obtained by standard resistance measurement, Figs.\ \ref{magres} (a) and (b), with the second derivative revealing oscillations in finer details.

The oscillatory part of $d\rho_{xx}/dB$ obtained by subtracting the slowly-varying background is shown in the main panel of Fig.\ \ref{subtract}. Here we employed an empirical function $\alpha \sqrt{B}$ that describes the slowly-varying part remarkably well for $|B|$$<$$\sim$25 mT as shown in Fig.\ \ref{rawdRdB}. The use of such a simple function as a background to be subtracted eliminates the possibility to inadvertently introduce artificial oscillations during the course of data processing. Oscillations can be discerned down to $\sim$2.5 mT in the figure. It is interesting to note the apparent resemblance of the oscillatory part with the CO observed at higher magnetic fields shown in the inset.

\begin{figure}
\includegraphics[bbllx=50,bblly=50,bburx=530,bbury=400,width=8.5cm]{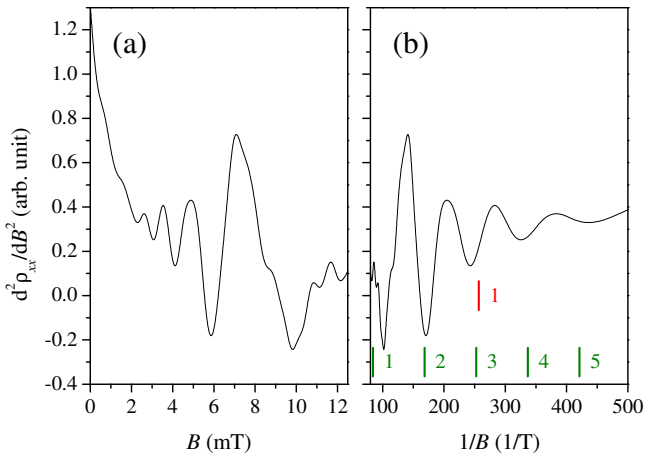}%
\caption{(a) The plot of $d^2\rho_{xx}/dB^2$ vs. $B$ (the close-up of the part enclosed by the box in Fig.\ \ref{magres}(d)). (b) Replot of $d^2\rho_{xx}/dB^2$ against $1/B$. The vertical lines with indices indicate the positions for the $n$-th geometric resonance given by Eq. (\ref{GROO}): $b_{0,1}$=$a$ (upper line) and $b_{1,2}$=$na$ with $n$=1, 2, 3, 4, 5 (lower lines). \label{d2dBBBI}}
\end{figure}

As in the previous papers \cite{Endo05N,Endo05I,Endo06E,Endo07I} and also shown in the inset of Fig.\ \ref{magres}, we use the notation $(j,k)$ to represent the open orbit generated by repeated alternating diffraction from the reciprocal lattice points $j q$ and $k q$ ($j$=0 signifies the absence of the corresponding diffraction) due to the $j$-th and $k$-th harmonic components $V_j$ and $V_k$, respectively, in the modulation potential. The width of the open orbit $(j,k)$ is given by
\begin{equation}
b_{j,k}=R_\mathrm{c} \left[ \sqrt{1-(j/\lambda)^2}- \sqrt{1-(k/\lambda)^2} \right],
\label{OOwidth}
\end{equation}
with $\lambda$=$a k_\mathrm{F}/\pi$ representing the number of minibands below the Fermi energy $E_\mathrm{F}$. The width $b_{j,k}$ inherits the inverse proportionality to $B$ characteristic of the cyclotron radius, as is evident from the inclusion of $R_\mathrm{c}$ as a factor in Eq. (\ref{OOwidth}).

The GROO takes place at the conditions
\begin{equation}
b_{j,k}=na \hspace{10mm}(n=1, 2, 3,...)
\label{GROO}
\end{equation}
and manifests itself as slight enhancement in the conductivity $\delta \sigma_{yy}$ hence in the resistivity $\delta \rho_{xx}$. Since $d\rho_{xx}/dB$ shown in Fig.\ \ref{rawdRdB} (or its oscillatory part in Fig.\ \ref{subtract}) is rather difficult to be directly related to the small enhancement in the resistivity, we instead make use of the second derivative $d^2\rho_{xx}/dB^2$ for analysis, whose minima have been shown to correspond to the local maxima in $\rho_{xx}$ \cite{Endo05N}. The second derivative obtained by numerically differentiating $d\rho_{xx}/dB$ in Fig.\ \ref{rawdRdB} is shown in Fig.\ \ref{d2dBBBI}(a), which is simply the expanded view of the part enclosed by a rectangle in Fig.\ \ref{magres}(d). Replot of the trace against $1/B$ shown in Fig.\ \ref{d2dBBBI}(b) reveals that the oscillations are periodic in $1/B$, as expected from Eqs.\ (\ref{OOwidth}) and (\ref{GROO}). As shown in the figure, the positions of minima in $d^2\rho_{xx}/dB^2$ (corresponding to the positions of maxima in $\rho_{xx}$) are found to take place at the locations expected from Eq.\ (\ref{GROO}): $b_{1,2}$=$na$ with $n$=1, 2, 3, 4, 5. The minimum at $B$=4.1 mT ($1/B$=243 T$^{-1}$) appears to explicable either by $b_{1,2}$=$3a$ or by $b_{0,1}$=$a$ equally well. In the previous papers, it was ascribed to the latter, since three minima at $B$=5.8, 4.1, and 3.1 mT ($1/B$=170, 243, and 327 T$^{-1}$) were not resolved and appeared as a single minimum because of the poor resolution [see also Fig.\ \ref{magres}(b)]. With the higher resolution, the former interpretation appears to be more natural, although the latter resonance may also be playing its share of role in the occurrence of the minimum.

Equation (\ref{GROO}) is analogous to the corresponding relation in CO, in which resistivity maxima take place at
\begin{equation}
2 R_\mathrm{c}=(n+1/4)a \hspace{10mm}(n=1, 2, 3,...).
\label{COpeak}
\end{equation}
Both Eqs.\ (\ref{GROO}) and (\ref{COpeak}) imply the oscillations periodic in $1/B$, which is the obvious reason for the apparent similarity between GROO and CO demonstrated in Fig.\ \ref{subtract}. The factor $[ \sqrt{1-(j/\lambda)^2}- \sqrt{1-(k/\lambda)^2} ]$ (the ratio of the width of the open orbits to the cyclotron radius) in Eq.\ (\ref{OOwidth}) for our present sample is $\sim$10$^{-2}$, which accounts for the order of magnitude difference between GROO and CO in the magnetic-field scale. Equation (\ref{COpeak}) includes the phase factor ``$1/4$'', while Eq.\ (\ref{GROO}) contains no counterpart. Equation (\ref{GROO}) naively expresses the matching of the open-orbit width with the integer multiple of the period (where we assume the enhancement in $\sigma_{yy}$), and is not necessarily based on firmly established theoretical background. Strictly speaking, therefore, the possibility that a certain phase factor should be included in Eq.\ (\ref{GROO}) cannot be completely ruled out. Our present result shown in Fig.\ \ref{d2dBBBI}(b), however, suggests that such phase factor is unnecessary.

The observation of GROO is inevitably limited to a rather narrow range of the magnetic fields \cite{Endo05N,Endo06E}; the magnetic field cannot be too small for the amplitude $\delta\rho_{xx} \propto B^2 \delta\sigma_{yy}$ to significantly exceed the noise level of the resistance measurement, while the upper limit is set by the magnetic-breakdown effect \cite{Streda90} from which the open orbit should survive to be able to be observed. In our present sample, the resonant conditions $n$=1, 2, ..., 5 in Eq.\ (\ref{GROO}) for the open orbit $(1,2)$ fall in the magnetic-field window ($\sim$2--50 mT), allowing the resultant $\delta\rho_{xx}$ to be observed. Although the resonant conditions with still higher $n$ for open orbits $(j,k)$ with $2\leq j < k$ also enter the magnetic-field window \footnote{The inset of Fig.\ \ref{magres} shows only three open orbits $(j,k)$ with $j < k \leq 3$. This is simply because $q$ is exaggerated compared with $k_\mathrm{F}$ in the figure for clarity. Open orbits with $k$ up to 8 is allowed for our present sample.}, these are less likely to be detected owing to the filtering action of the magnetic-breakdown effect; the effect favors the open orbit $(j,k)$ with smaller values of $j$ and $k$, since diffractions caused by higher harmonic components of the modulation potential ($V_j$ and $V_k$ with larger values of $j$ and $k$, generally having smaller amplitudes) possess higher probability of the magnetic-breakdown \cite{Endo06E}. They may, however, be resolved with further improvement in the magnetic-field resolution; slight deviation seen in Fig.\ \ref{d2dBBBI}(b) of the positions of minima from the positions expected from Eq.\ (\ref{GROO}) may possibly be caused by the intervention of such higher-order terms of open orbits involving higher harmonic potential contents. For more quantitative explanation of the oscillation amplitudes of the observed GROO, a theory that quantitatively explains the enhancement in $\sigma_{yy}$ by the resonance is required, which remains to be explored. 

To summarize, we have observed the geometric resonance of open orbits (GROO) down to very low magnetic fields ($\sim$ 2.5 mT) by improving both the magnetic-field resolution and the sensitivity in the resistance measurement. The resonance reveals clear periodicity in $1/B$, with the positions of the resistivity maxima (minima in the second derivative) described well by Eq.\ (\ref{GROO}) with $n$ up to 5, precluding the necessity of an additional phase factor.

\ack
This work was supported by Grant-in-Aid for Scientific Research (C) (18540312) and (A) (18204029) from the Ministry of Education, Culture, Sports, Science and Technology.










\bibliography{lsls,ourpps,twodeg}

\end{document}